\newcommand{\ket}[1]{{|#1\rangle}}
\newcommand{\calC}{{\cal C}}
\newcommand{\e}{\epsilon}

\documentclass{article}
\usepackage{graphics}
\begin{document} 
\title{Threshold Estimate for Fault Tolerant Quantum Computing}
\author{Christof Zalka \\ e-mail \quad zalka@t6-serv.lanl.gov}
\maketitle

\begin{abstract}
I make a rough estimate of the accuracy threshold for fault tolerant quantum
computing with concatenated codes. First I consider only gate errors and use
the depolarizing channel error model. I will follow P.Shor \cite{shor} for
fault tolerant error correction (FTEC) and the fault tolerant implementation
of elementary operations on states encoded by the 7-qubit code. A simple
computer simulation suggests a threshold for gate errors of the order
$\e\approx 10^{-3}$ or better. I also give a simple argument that
the threshold for memory errors is about 10 times smaller, thus $\e\approx
10^{-4}$.
\end{abstract}

\section{Introduction}

In the past 2 years ever better results have been achieved concerning what in
principle can be done with the help of quantum error correcting codes. For a
very readable review see e.g. \cite{preskill1} (\cite{preskill2} by the same
author is more generally about quantum computers). In particular it has been show
\cite{knill,aharonov} that once elementary unitary operations (``gates'') in a
quantum computer can be carried out with more than some threshold accuracy and
also decoherence on resting qubits is sufficiently low, then it is possible to
carry out arbitrary precision operations on suitably encoded ``computational''
qubits. From this it follows that one can then carry out computations with any
number of steps without errors accumulating too much. 
In this paper I will take a more practical, engineer-like point of view. I
will follow P.Shor \cite{shor} for fault tolerant error correction (FTEC) and
the fault tolerant implementation of elementary operations on states encoded
by the 7-bit code. I consider only non-systematic stochastic gate errors and
assume that gates can be applied to any two (or three) qubits of the quantum
computer. Also I will use (and try to justify) the most simple and natural
error model. With these assumption, computer simulation of an optimized
version of Shors techniques indicates an (astonishingly high) threshold of the
order $\e \approx 10^{-3}$ on the tolerable error probability per gate. I will
give arguments why the threshold can actually be expected to be even better
(=higher) than that. For comparison I also provide a very rough calculation by
hand.

The simulation I actually make to get the mentioned result is very simple. I
consider only one level of encoding and try to find the error rate below which
the encoding starts to pay off. Also I look only at one encoded qubit to
which 1-qubit operations are applied.

\section{On the error model}

I would like to sketch a justification for the simple error model I use. The
model is as follows: First of all I assume that errors happen with a certain
probability (per gate). I also assume that these probabilities are
independent for different qubits, except when an operation acts on several (2
to 3) qubits at once (like an XOR). For 1-bit errors I assume that the three
possible errors (bit-flip, phase-flip and combined bit and phase-flip) occur
all with probability $\frac{1}{3}\e$ (depolarizing channel). This is
very natural as it is equivalent to the admixture of the unit density matrix
to the original pure state. The analogous assumption will be made for several
qubits taking part in an operation. Thus the 15 different errors of the two
qubits which took part in an XOR are equally probable. The size is chosen
such, that when we look only at one of the two qubits, we get the same
probabilities as for a 1-qubit error. Actually one sees that the quantity
$\frac{4}{3}\e$ would be more natural than the usual $\e$. For simplicity I
also assume that the error probability is the same for all operations and that
such errors are the only ones occurring. Thus I e.g. don't consider
decoherence on ``resting'' qubits.

The motivation for only considering gate errors is as follows: It is
imaginable that the decoherence on resting qubits can be made very small. On
the other hand gates, in some sense being analog, will always have some error
in their continuous parameters. Such parameters are e.g. given by the
intensity and time of a laser beam shining at an ion in an ion trap quantum
computer. Also, the interaction with exterior fields during the gate will cause
additional decoherence.

It would actually be easy to adapt my computer program to the case where
decoherence on resting qubits is important. But then the result will depend on
the degree of parallelism with which the quantum computation can be carried
out and this in turn depends on the physical realisation of the QC. E.g. in an
ion trap QC with just one center of mass motion bus-qubit, hardly any
parallelism is possible. If different phonon modes could be used, some
parallelism would be possible. 

Nevertheless I have added a small paragraph roughly estimating the threshold
for memory errors under a reasonable, if somewhat arbitrary, assumption on
parallelism.

For {\bf decoherence} a probabilistic error model immediately seems
reasonable, as a mixed state (which decoherence produces) can be
viewed as an ensemble of pure states together with their probabilities
of occurrence. For this to be possible, the ensemble of states, taken as
a basis, must diagonalize the density matrix. For depolarizing channel
type of decoherence (which we assume here) this is true for any
orthogonal set of states which includes the undisturbed original
state.

A second source of errors are {\bf gate inaccuracies} (unitary errors), thus
deviations of the parameters of the unitary operation from the desired
value. Let's imagine a whole ensemble of QCs carrying out the same operation
on the same state and assume that the unitary errors in different QCs are
independent and have average 0 (=statistical errors, as opposed to systematic
ones). Provided we know nothing about the actual errors in the individual QCs,
the resulting ensemble of states can at best be described by a density
matrix. For the appropriate natural assumption on the distribution of the
unitary errors, this leads to the depolarizing channel error model. The
``natural'' assumption is obviously that the error probability is
``isotropic'', thus, loosely speaking, that it is the same for any three (for
SU(2)) orthonormal parameters.

Quite another problem are {\bf systematic unitary errors} (which we neglect
here), e.g. when a rotation on a qubit tends to overrotate in all QCs. Such
errors could e.g. be diminished by some feedback mechanism which would correct
gates after several preliminary test runs. Also it can be expected that such
errors would to some degree average each other out in some given quantum
computation, contrary to the simple accumulation that would take place if we
simply applied a gate a number of times to the same qubit. In this later case
the effective error rate would roughly equal the amplitude error instead of
its square, as for non-systematic errors.

\section{Review of Shors fault tolerant error correction}

\subsection{The 7-bit code}

The codes used by Shor are of the type described by Calderbank and Shor
\cite{calder} and Steane \cite{steane}, where a quantum code is constructed
from two (classical) binary linear codes $\calC_1$ and $\calC_2$ with
$\calC_2^\perp \subset \calC_1$. For Shors code we have $\calC_1=\calC$ and
$\calC_2=\calC^\perp$, thus $\calC^\perp \subset \calC$ and also
$dim(\calC)-dim(\calC^\perp)=1$, thus these codes encode 1 qubit. Such a code
can be obtained from a self dual ($\calC=\calC^\perp$) Reed-Muller code by
leaving away one of the $2^m$ bits. For the threshold the smallest such code
that still can correct for 1 error is relevant. This is the 7-bit code given
by the following 4 basis elements:
\\ 
\\1011100  
\\1101010  
\\1110001 
\\0100011
\\ \\
The dual code $\calC^\perp$ is given by the first 3 of these elements, thus
$l=dim(\calC)=4,~ dim(\calC^\perp)=3$, and the length $n=7$. In the so 
called ``s-basis'' the quantum codewords are
\begin{displaymath}
\ket{s_v}=2^{-(n-l)/2} \sum_{w\in \calC^\perp} \ket{w+v}
\qquad \mbox{where} \quad v \in \calC
\end{displaymath}
Actually there are only 2 different codewords, and
they will represent an encoded 0 resp. 1:
\begin{displaymath}
\ket{0_L}=\ket{s_v} \qquad \mbox{for any} \quad v \in \calC^\perp
\end{displaymath}
and
\begin{displaymath}
\ket{1_L}=\ket{s_v} \qquad \mbox{for any} \quad v \in \calC / \calC^\perp 
\quad (\mbox{thus}~ v \in \calC,~ \mbox{but}~ v \not \in \calC^\perp).
\end{displaymath}
Where $L$ stands for ``logical'' which means the same as ``computational''.
By applying a Hadamard transformation to each of the 7 qubits we go to the
``c-basis''. One can show that:
\begin{displaymath}
\ket{c_v}=H^7~ \ket{s_v}=2^{-l/2} \sum_{w \in \calC} (-1)^{v\cdot w}~ \ket{w}
\end{displaymath}

Thus $\ket{c_v}$, like $\ket{s_v}$, is a linear combination of codewords
out of $\calC$.
\subsection{Syndrom measurement}

In the theory of classical linear codes the syndrom of a received (and
possibly distorted) codeword is obtained by scalar multiplication with 3 basis
elements of $\calC^\perp$. Clearly when all 3 syndrom bits are 0 then there is
no error in the codeword. The other 7 possibilities each correspond to a
particular one of the 7 bits having flipped. Thus knowledge of the syndrom
allows to restore the original codeword (provided there is at most 1
error). For quantum error correction it is crucial that one can measure the
syndrom without measuring the encoded qubit. Encoded qubits $\alpha
\ket{0_L}+\beta \ket{1_L}$ are in either basis linear combinations of basis
states $\ket{v}$ with $v \in \calC$. A bit-flip in one qubit will cause the
same syndrom in all $\ket{v}$'s, thus measuring the syndrom will not collapse
the encoded qubit. Phase-flips don't affect the syndrom and combined bit and
phase-flips will look like bit-flips. Now a Hadamard transform on a qubit will
change a bit-flip into a phase-flip and vice versa. Thus by measuring the
syndrom in both, the $s$- and the $c$-basis we get the full ``quantum
syndrom'' which allows us to restore the original state by applying
appropriate flips to the state. We see here that the 7-bit code can actually
handle more than just 1-bit errors as long as there is only one bit-flip and
one phase-flip (combined bit and phase-flips count as both).

To measure the syndrom bit corresponding to some $v \in \calC^\perp$, we could
take an auxiliary qubit in the initial state $\ket{0}$ and then XOR all qubits
in the encoded state at positions where $v$ has a 1 into it. Note that all
non-zero elements of $\calC^\perp$ have four 1's.

\subsection{Fault tolerant syndrom measurement}

Shor proposes to measure syndrom bits in a different way. For every such
measurement we need an auxiliary 4-qubit ``cat'' state 
\begin{displaymath}
\ket{cat}=\frac{1}{\sqrt{2}} (\ket{0000}+\ket{1111})
\end{displaymath}
which we Hadamard transform to get
\begin{displaymath}
H^4~ \ket{cat}= 2^{-3/2} \sum_{|x|~ even} \ket{x}
\end{displaymath}
thus a linear combination of all 4-bit words with an even number of 1's. Now
instead of XORing the 4 qubits of the codeword into a single auxiliary qubit,
we XOR each one into one of the 4 auxiliary qubits. Then we observe this
state. The parity bit of the observed 4-bit word will then be the syndrom
bit. One can wonder whether we collapse the encoded state in an unwanted way
by observing 4 qubits instead of just the syndrom bit, but one can show that
this is not so.

\subsection{What is fault tolerance?}

If computation with encoded qubits is to pay off, the error probability (of
making a codeword uncorrectable), should be smaller than the fundamental error
rate $\e$ one would have without encoding. This can be achieved by making sure
that an error in error correction affects at most one bit of the codeword. Then
the probability of introducing errors into several qubits and thus making the
codeword uncorrectable is of order $\e^2$. Thus by making $\e$ small enough we
can make sure that encoding pays off. A conventional syndrom measurement with
just 1 auxiliary qubit isn't fault tolerant, because, as one can check, the
probability of affecting several qubits is of order $\e$, not $\e^2$ as with
Shors method.

\section{Fault tolerant quantum computation}

\subsection{Operations on encoded qubits and FTEC}

To achieve fault tolerant quantum computation, besides fault tolerant error
correction (FTEC) we need a way to apply computational operations to the
encoded qubits such that again the error probability is of order $\e^2$. For
the 7-bit code several operations can be carried out bitwise, that is, to
apply an operation to an encoded qubit we have to apply it individually to all
7 qubits. This is clearly fault tolerant. Pauli matrix operations (Clifford
group), NOT, phase flips and the XOR between 2 computational qubits can be
carried out in this way. To be able to carry out any calculation, we need an
additional operation, e.g. the Toffoli gate. Applying a Toffoli gate to three
encoded qubits is not this simple, but it can also be done fault tolerantly
(see \cite{shor}). In fault tolerant quantum computation we carry out
operations on the encoded computational qubits and once in a while we apply a
recovery operation to prevent the acculmulating correctable errors from
becoming uncorrectable ones. For fault tolerant quantum computation we also
need to be able to prepare e.g. an encoded $\ket{0}$ and to observe encoded
qubits:

\subsection{Making $\ket{0_L}$ and observing encoded qubits}

For preparing encoded states $\ket{0_L}$ I propose to start with an arbitrary
7-qubit state and then essentially apply FTEC to it till we get $\ket{0_L}$,
where the possibility to throw away states that don't seem to have come out
right, is important. To be able to discriminate between $\ket{0_L}$ and
$\ket{1_L}$, we need ``syndrom'' bits other than $v \in \calC^\perp$, namely
we need at least one $v \in \calC/\calC^\perp$. We could proceed as follows:
We take an arbitrary 7-qubit state and apply error correction to it. In
addition to the usual $2\times 3=6$ syndrom bits we have to measure a
``0/1-syndrom bit'' telling us whether we have $\ket{0_L}$ or $\ket{1_L}$. If
we get $\ket{1_L}$ we can either throw it away or we can change it to
$\ket{0_L}$ by negating all 7 qubits. Then we have to {\bf verify} the thus
obtained state, e.g. by again measuring all 6+1 syndrom bits. If they are not
all 0 we start over. The probability of getting a $\ket{1_L}$ instead of a 
$\ket{0_L}$ then is of order $\e^2$, as required.

The observation of an encoded qubit should be done in the $s_v$ basis, as only
there one can tell $\ket{0_L}$ from $\ket{1_L}$ by observing a 7-bit word
belonging to the superposition. Because these words are in $\calC$ we get an
error probability of $O(\e^2)$.

\section{My (improved) implementation of Shors FTEC}

Here I describe the improved fault tolerant error correction which I have
simulated on a computer. First a cat state has to be constructed. This can be
done by resetting 4 qubits to $\ket{0}$ and then Hadamard transform the first
one to get $1/\sqrt{2}~(\ket{0}+\ket{1})\otimes \ket{000}$. Then I apply an
XOR from the 1. to the 2. qubit then from 2. $\to$ 3. and finally from
3. $\to$ 4. Now this state would have a probability of order $\e$ to have more
than 1 bit-flip, which would be harmful. Thus, as Shor points out, we have to
verify this state. I take an additional auxiliary qubit in state $\ket{0}$ and
XOR the 1. and the 4. cat qubit into it. If upon observation this qubit isn't
0 we have to try again to construct a cat state. Hadamard transforming the 4
qubits in the last step is of course fault tolerant.

\subsection{Syndrom measuring strategies}

As Shor also points out, it is not enough to measure the syndrom just once to
get error probability $O(\e^2)$. On the other hand we should minimize the
number of syndrom bit measurements as they threaten to destroy the encoded
qubit. Instead of repeating the whole syndrom measurement (in one basis, say
the $s$-basis), I measure the parity bit of the 3 just measured syndrom
bits. Thus if $v_1,v_2,v_3$ are the 3 basis elements of $\calC^\perp$, I next
measure the syndrom bit corresponding to $v_4=v_1+v_2+v_3$. If this parity
check turns out to be wrong, I again measure the parity bit of the last three
measurements, thus $v_5=v_2+v_3+(v_1+v_2+v_3)=v_1$, and so on till the last 4
measurements are consistent. This strategy has the potential advantage that
errors which have been introduced into the codeword by early syndrom bit
measurements, may be detected and corrected. Actually there is one case in
this sceme where a FTEC step can destroy an originally error free encoded
qubit with probability $O(\e)$. This is when the last two measured syndrom
bits are 1 and the two preceeding ones are 0. This case can specifically be
taken care of.

Another improvement can be made when the syndrom indicates an undisturbed
codeword, as will be the case most of the time. In this case I don't measure
the syndrom parity bit, as no erroneous error correction attempt threatens to
introduce an additional error into the codeword. Actually we can even go
further and abandon the error correction step if the first syndrom bit
measurement yields 0. Of course, which one we take as the first one should
then cyclicly be changed between $v_1, v_2, v_3$ from error correction step to
error correction step. In a way this is like carrying out only $1/3$ of an
error correction step, but as actually 4 out of 7 errors will show up in a
given single syndrom bit, it pays off. I will refer to this sceme as ``1/3
FTEC''.

\subsection{Reducing the number of Hadamard transforms}

Schematically a syndrom bit measurement in the $s$-basis (thus looking for
bit-flips) goes as follows:

\begin{picture}(350,100)

\put(10,67){$\ket{\psi}$}
\put(30,70){\vector(1,0){270}} 

\put(0,27){$\ket{cat}$}
\put(30,30){\vector(1,0){50}} \put(85,27){$H$}
\put(100,30){\vector(1,0){180}} \put(285,27){$M$}

\put(150,68){\vector(0,-1){36}} \put(120,47){$XOR$}

\end{picture}
where $H$ stands for a Hadamard transformation on each qubit and $M$ means
measurement. The XOR applys only to the 4 qubits of the codeword
$\ket{\psi}$ which are at the positions of 1s in $v \in \calC^\perp$.

Measuring a single syndrom bit in the $c$-basis is more complicated, as we
first have to transform the codeword to the $c$-basis and then back again:

\begin{picture}(350,100)

\put(10,67){$\ket{\psi}$}
\put(30,70){\vector(1,0){50}} \put(85,67){$H$}
\put(100,70){\vector(1,0){100}} \put(205,67){$H$}
\put(220,70){\vector(1,0){80}}

\put(0,27){$\ket{cat}$}
\put(30,30){\vector(1,0){50}} \put(85,27){$H$}
\put(100,30){\vector(1,0){180}} \put(285,27){$M$}

\put(150,68){\vector(0,-1){36}} \put(120,47){$XOR$}

\end{picture}
But an XOR conjugated with Hadamard transforms is simply an XOR in the
opposite direction, thus we get the same result with:

\begin{picture}(350,100)

\put(10,67){$\ket{\psi}$}
\put(30,70){\vector(1,0){270}} 

\put(0,27){$\ket{cat}$}
\put(30,30){\vector(1,0){170}} \put(205,27){$H$}
\put(220,30){\vector(1,0){60}} \put(285,27){$M$}

\put(150,32){\vector(0,1){36}} \put(120,47){$XOR$}

\end{picture}
\section{Iterated encoding (concatenated codes)}

If the error per operation is smaller on the encoded level than on the
fundamental level, it seems that by iterated encoding we can achieve an
arbitrary precision. So on the next level we would encode codewords with the
7-bit code where each qubit would again be encoded with this code (thus 49
qubits per computational qubit). This requires that all operations needed for
FTEC can also be carried out on the encoded level (with the reduced error
probability). The necessary operations are: Hadamard transformation, XOR,
preparation of $\ket{0}$ and observation of a qubit. For my threshold result
it will be crucial that the Toffoli gate isn't needed for FTEC. The Hadamard
transform and the XOR can be carried out bitwise (transversally), thus they
can be carried out with the effective error rate
$\e_1~\stackrel{\textstyle <}{\sim}~ 10^3~ \e_0^2~$, where $\e_0=\e$.

\section{How to estimate the threshold}

\subsection{simulating errors without knowing the state}

With the depolarizing channel error model it is possible to very efficiently
simulate the errors and their propagation without actually having to store the
potentially hughe quantum state of the QC. All we need are two classical bits
for each qubit in the QC. One bit indicates whether a bit-flip has happened
while the other one keeps track of phase-flips. When a Hadamard transformation
is applied to a qubit the bit-flip and phase-flip bits have to be
exchanged. An XOR between two qubits adds the bit-flip of the source qubit to
the target qubit while phase-flips propagate in the opposite direction.

The computer simulation I do is really a Monte Carlo simulation. Thus I run
the same ``quantum computation'' many times and introduce at each gate errors
with some probability.

\subsection{a (strongly) simplified ``quantum computation''}

What quantum computation do I actually simulate? I consider a single encoded
qubit (taking up $7\times 2$ classical bits in my program). I start with an
error-free codeword, then I apply alternatingly a number of computational
1-qubit operations and a FTEC step to it. I assume all 1-qubit operations to
be applicable bitwise. Thus for my simulation this simply means, that I
introduce errors into the individual qubits with the appropriate
probability. The FTEC step is more elaborate, as I have to construct cats,
measure syndrom bits with them,... . It turns out that there is a optimal
number of 1-qubit operations between error correction steps. More generally
there is a optimal probability of 1-qubit errors in the codeword when FTEC
should be applied. This is because the probability that the computational
operations produce an unrecoverable error in the codeword increases
quadratically with the number of operations. On the other hand FTEC should
also not be applied too often, as there is some probability (of course of
order $\e^2$) that it destroys even an error-free codeword. The optimal number
of 1-qubit operations before a ``1/3 FTEC'' step turns out to be about 5, for
a full FTEC step it's about 15.

\subsection{the computer simulation}

In the simulation, I now go on applying a number of 1-qubit operations followed
by an FTEC step to the same codeword until it gets an uncorrectable error. I
then calculate the average number of 1-qubit operations I can carry out like
that. The inverse of this gives the effective error probability for the
encoded qubit per 1-qubit operation. A first rough estimate of the threshold
is the error rate for which this effective error rate is equal to the
fundamental error rate.

I provide two figures with data from the computer simulation. For clarity I
used only ``full FTEC'' steps for both figures even though with ``1/3 FTEC''
steps somewhat better results could be obtained. 

The first figure
shows how the error rate per 1-qubit gate on the encoded level changes with
the number of 1-qubit gates applied to the encoded qubit between full FTEC
steps. So this error rate is the error rate per FTEC step divided by the
number of 1-qubit operations between FTEC steps. The function is of the form
$a/x+b+c x$. The first term comes from the probability of an FTEC step to
destroy an undisturbed codeword (destroy=introduce an uncorrectable
error). The second (constant) term comes from a probability of destroying a
codeword in an FTEC step which is proportional to the probability of
correctable errors in the codeword before the FTEC step. The last term is the
probability that the codeword gets an uncorrectable error by accumulating
1-qubit errors from operations, thus this term is independent of the FTEC
procedure. (For these three terms see also the paragraph about hand
calculations.) From the data one can see that around the optimal number of
operations between FTEC steps, which is about 15, the middle (constant) term
dominates. Thus when trying to improve my FTEC procedure, this term should be
targeted. It is also interesting to note that the error rate per operation
doesn't increase much if we choose to make FTEC steps less often (e.g. to save
time).
\vspace*{1cm}
\resizebox{13cm}{9cm}{\includegraphics[18,144][588,566]{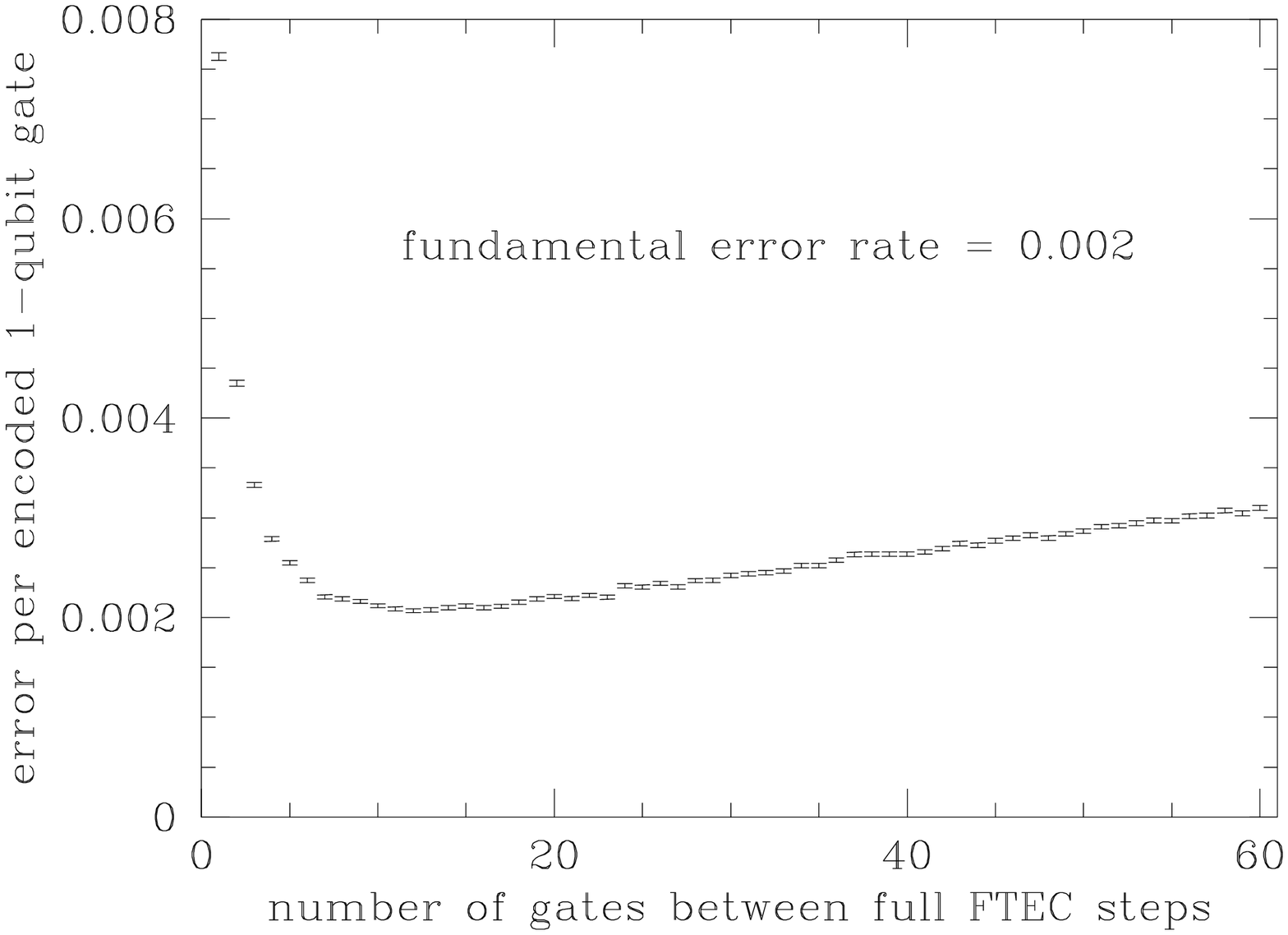}} \\
\vspace*{1cm}

The second figure shows how the error on the encoded level per 1-qubit gate
applied to the encoded qubit increases with the fundamental error rate. The
number of operations between full FTEC steps has been set to the optimal
number of 15. Fault tolerance, as I have defined it, means that the error rate
on the encoded level is quadratic in leading order in the fundamental error
rate. The deviation from a parabola can be seen in the right half of the
figure. The ``break-even'' error rate can be seem to be about 0.002 . I claim
that this ``break-even'' error rate is a reasonably good estimate of the
threshold error rate (here only for gate errors).

\resizebox{13cm}{9cm}{\includegraphics[18,144][588,566]{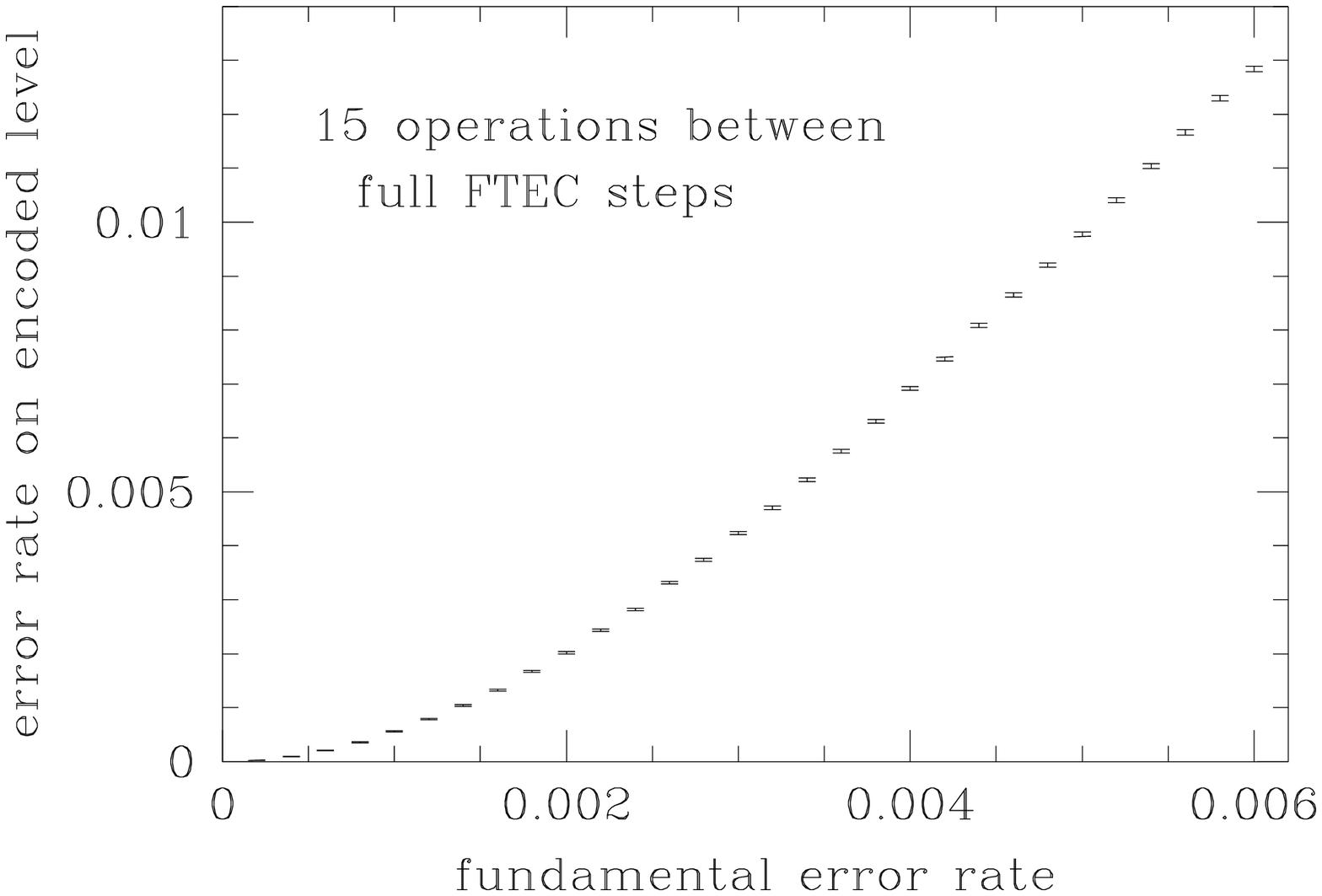}}

Following is a listing of the C-program that produced the data in the 2
figures. I ran it on a linux machine with gcc.

\begin{verbatim}

#include <stdio.h>
#include <math.h>

#define Rep 10000
#define For(var,lim) for(var=0;var<(lim);var++)

double P=4.0/3*0.002;
int Nop=15;
int state[8][2], cat[4][2], ab[2];    /* total 12 qubits */
int code[6][7], synd[5];
int err[2][2][2];
int i1,ssum,dpr=0,dc3,db,*v; FILE *data;

double drand48(); void srand48(long);
int rb() {return drand48() < 0.5;}

void e1b(int *qb) {int i2; if(drand48()<P) For(i2,2) qb[i2]^=rb();}
void e2b(int *qb1,int *qb2) {int i2;
  if(drand48()<P) For(i2,2) {qb1[i2]^=rb(); qb2[i2]^=rb();}} 

void xor(int *qb1,int *qb2) {qb2[0]^=qb1[0]; qb1[1]^=qb2[1]; e2b(qb1,qb2);}

void Htr(int *qb) {int h; e1b(qb); h=qb[0]; qb[0]=qb[1]; qb[1]=h;}
void Hcat() {int k; For(k,4) Htr(cat[k]);}

void estate(float p) {int i2,j; 
       For(j,7) if(drand48()<p) For(i2,2) state[j][i2] ^=rb();}

void stab() {int i,j,i2,z,z1; 
For(i2,2) {For(i,3) {z=0; 
  For(j,7) z^=code[i][j]*state[j][i2]; synd[i]=z;}
  z=err[synd[0]][synd[1]][synd[2]];
  z1=0; For(j,7) z1^=state[j][i2]; 
  For(j,7) state[j][i2]=0; if(z != 7) state[z][i2]=1;
  if(z1 == (z==7)) state[7][i2]^=1;}}

void makecat(float p,int n) { int k,h;
do{For(k,n) {cat[k][0]=cat[k][1]=0; e1b(cat[k]);}
   e1b(cat[0]);
  for (k=1; k<n; k++) xor(cat[k-1],cat[k]);
  ab[0]=ab[1]=0; e1b(ab);
  xor(cat[0],ab); xor(cat[n-1],ab);
  e1b(ab);} while(ab[0] != 0);}

void makecode() {int c[6],i,j; 
c[0]=1110100;
c[1]=1001110;
c[2]=1101001;
c[3]=1010011;
c[4]=1111111;
c[5]=1100010; For(i,6) For(j,7) {code[i][6-j]=c[i]%10; c[i]/=10;}}

void prstate() {int j,i2;
For(i2,2) {For(j,7) printf("%d ",state[j][i2]); 
  printf("  %d\n",state[7][i2]);} printf("\n");}

int syndv(int n,int i2) {int j,k=0,z; makecat(P,n);
if(i2==0) {Hcat(); For(j,7) if(v[j]){xor(state[j],cat[k]); k++;}}
if(i2==1) {For(j,7) if(v[j]){xor(cat[k],state[j]); k++;} Hcat();}
z=0; For(k,n) {e1b(cat[k]); z^=cat[k][0];} return z;}

int synd01() {v=code[5]; return syndv(3,0);}
    
void ftec(int i2) {int i,j,k,z;
for(i=0; ; i++) {if(db) printf("%d",i);
   v=code[(i+i1)%4]; synd[(i+i1)%4]=syndv(4,i2);  
  ssum=synd[0]^synd[1]^synd[2]^synd[3];
  if(synd[(i+i1)%4] && !synd[(i+i1+1)%4] && 
         !synd[(i+i1+2)%4] && synd[(i+i1+3)%4]) ssum=1;
  /* if( i==0 && synd[i1%4]==0) {synd[0]=synd[1]=synd[2]=0; break;} */
  if (i>=3 && ssum==0) break;}
    z=err[synd[0]][synd[1]][synd[2]];
    if(z!=7) {state[z][i2]^=1; e1b(state[z]);}
  if(db) printf("  %d \n",z);}

main(){ int i,i2,i3,j,k,z,bh; double dm,dv,sd,sdr,eps,eps1; 
srand48(time(NULL)); makecode(); err[0][0][0]=7;
For(j,7) err[code[0][j]][code[1][j]][code[2][j]]=j;

/* dv=0; For(i,10) {dv+=1000000000000000000LL; printf("%Ld\n",dv);} exit(0); */

/* data=fopen("gr","w"); for(Nop=1;Nop<=60;Nop++){ printf("%d\n\n",Nop); */
 data=fopen("grv","w"); for(i=1;i<=60;i++){eps=0.0002*i; P=4.0/3*eps; 
for(i3=0,dm=dv=0;;i3++){sdr=sd=1;
  For(j,8) {state[j][0]=state[j][1]=0;}
  for(i1=1;;i1++){dpr++; db=(dpr%Rep==0);
    estate(1-pow(1-P,Nop)); For(i2,2) ftec(i2);

stab();
if(db){prstate(); eps1=1.0*i3/dm; sdr=sqrt(1.0*i3*dv/dm/dm-1.0)/sqrt(i3); 
 sd=eps1*sdr; 
 printf("%d %f %d %1.9f %1.9f %1.5f\n\n\n",Nop,eps,i3,eps1,sd,sdr);}
  
if(state[7][0]+state[7][1]) {dm+=1.0*Nop*i1; dv+=1.0*Nop*Nop*i1*i1; break;} }
  if(sdr*sqrt(eps1/0.002)<0.02) {
  fprintf(data,"%f %f %1.7f\n",eps,eps1,sd); break;}
} }    fclose(data);


}

\end{verbatim}

\subsection{shortcomings}

First of all we have not considered 2- or 3-qubit gates. These will propagate
the 1-qubit errors in the codewords to other codewords and will thus make it
necessary to apply FTEC more often. To be on the save side, we could apply
full FTEC right after each gate in which a codeword has participated. For this
case the simulation gives a threshold of $\e \approx 1/2000$. But this much
FTEC will hardly be necessary. Say we apply just a ``1/3 FTEC'' step after
each gate. Then the threshold will be $\e \approx 1/700$.

Another simplification is the assumption that the error will diminish in the
same way from one level of encoding to the next, as it does from the
fundamental level to the first one. This is not true. Actually also the error
pattern changes so that we should really consider more that just the
probability $\e$ to describe it. There are reasons to expect the error
evolution to higher levels to become worse ``higher up'' but there are also
good arguments to the contrary (see below).

\subsection{The Toffoli gate}

So far I have only considered gates that can be applied bitwise to encoded
qubits. This is not true for the Toffoli gate, although Shor \cite{shor} gives
a (complicated) fault tolerant implementation of the Toffoli gate on encoded
qubits. I claim that nevertheless the Toffoli error will go down with
additional levels of encoding as soon as this is true for the errors of the
simpler operations. This is mostly because the Toffoli gate isn't needed for
FTEC. A very rough estimate of the coefficients gives the following evolution
equation for going from level $l$ to level $l+1$ for the Toffoli error:

\begin{displaymath}
\e_{T,l+1}\approx 5~ \e_{l+1}+10~\e_{T,l}~ \e_l+20~{\e_{T,l}}^2~ \e_l
 +\frac{1}{512}~ {\e_{T,l}}^3 
\end{displaymath}
The implementation of the Toffoli gate consists mostly of the preparation of a
complicated auxiliary state. That's also where the next lower level Toffoli
gate has to be used. Because we can verify the integrity of this
state and redo it if necessary, there is no term $\e_{T,l}^2$ in the above
equation. The smallness of the last 2 coefficients is mainly due to my
possibly too benign error
model where an error in a Toffoli gate is likely to affect all 3 bits, and
thus is easily detected. But one sees that even for much larger
coefficients the Toffoli error will be ``dragged down'' once $\e_l$ diminishes.

\subsection{what counts is that FTEC gets ``better'' on higher levels}

Actually the same argument about ``dragging down'' can be applied to other
than the Toffoli gate errors. The statement is that what really counts is that
higher-level FTEC can be improved because once this can be achieved, gate
errors can be taken care of by applying FTEC often enough. So for the
threshold it is enough to analyze the ``inner workings'' of a higher-level
FTEC. Most of these ``inner workings'' are XORs. Say we apply a ``1/3 FTEC''
step on the two encoded qubits before applying the XOR. One can check that
this more than counteracts the increase of errors in the XOR due to new errors
and due to ``copying'' errors (error propagation). Thus this procedure should
keep the frequency of lower-level errors under control. I think this should
justify the assumptions which led to the above estimate of $\e \approx 1/700$.

Let's make a rough little calculation concerning the application of ``1/3
FTEC'' on both encoded qubits before each XOR. Let $x$ denote the error
probability per qubit in units of the fundamental error probability $\e$. Then
the ``1/3 FTEC'' step will reduce this to $3/7 ~x$. The XOR will increase the
error probability by $50\%$ due to error copying (error propagation) and
introduce its own errors. Thus:

\begin{displaymath} x \to 3/7 ~x \to 1.5(3/7 ~x)+1 \end{displaymath}
This leads to the well acceptable equilibrium value $x=14/5$. Above we have
not considered errors introduced by the ``1/3 FTEC'' step and neither have
differentiated between bit-flips and phase-flips. With $x_b$ and $x_p$ for
those probabilities, a more accurate calculation is straight forward.

\subsection{why I expect the threshold to be larger}

There are reasons why the error may go down faster on higher levels than what
one might at first expect. First of all on higher levels auxiliary states can
be verified more thoroughly before they are used. E.g. a cat on the first
level consists of 4 codewords whose integrity we can check.

Then, as mentioned above, it is generally (not only for the Toffoli gate) true
that what really counts is whether we can improve the FTEC step when going to
higher levels. Thus we can deal with the error propagation problem
specifically for the sequence of operations that make up an FTEC step and can
try to optimize this.

\subsection{the threshold for ``memory errors''}

So far I have assumed that the errors introduced by \underline{operations} on
the qubits dominate. Let's now consider errors which affect all qubits at
every timestep (duration of an operation) irrespective of whether they take
part in an operation or not. It is clear that in the limit of large QCs, for
such errors a (non-zero) threshold only exist if we allow operations to be
carried out in parallel on different parts of the quantum computer. So this
threshold will critically depend on how much parallelism we assume QCs will be
capable of. Here I assume that for every 7-qubit codeword in the QC we have an
independent machinery. This machinery includes 4 auxiliary qubits for making
cats plus 1 qubit for verifying them. Within every such $12$-qubit block
I assume that all operations are carried out sequentially. 

The preparation and verification of a 4-qubit cat takes 12 operations. The
measurement of a single syndrom bit with this cat takes another 12 steps. Thus
while the 12 qubits on average are involved in 2 operations they each also
have to wait for 24 time steps. Thus on average 12 times more memory errors
than operational errors affect each qubit. From this we can state that the
memory-error threshold is roughly 10 times lower than the gate-error
threshold (for our assumptions on the degree of parallelism). 

I think that this type of rough estimates is adequate for the time being, as
we are far from an optimal fault tolerant quantum computation sceme and as we
know very little about the errors and the possible parallelism of a real
quantum computer.

\section{Calculation by hand}

Consider the following diagram representing $n$ bitwise 1-qubit operations in
a row, followed by an error correction step. Note that the $n$ operations
simply correspond to an error probability of $n\cdot \e$ (for $\e \ll 1$) per
qubit. Practically the error probability $n\cdot\e$ will be produced by fewer
than $n$ 1-, 2- or 3-qubit operations due to the propagation of lower level
errors, as mentioned above.

\begin{picture}(350,120)

\put(120,90){$n$ operations} \put(230,90){error correction} 

\put(-30,75){0 errors}                  \put(95,80){.}
\put(-30,45){1 error (correctable)}     \put(95,50){.}
\put(-30,15){2 errors (uncorrectable)}  \put(95,20){.}

\put(105,80){\vector(3,-2){90}} \put(150,68){$7 n \e$} \put(202,80){.}
\put(105,80){\vector(3,-1){90}}                        \put(202,50){.}
   \put(125,25){$\left({7\atop 2}\right) (n\e)^2$}     \put(202,20){.}

\put(210,80){\vector(3,-2){90}} \put(255,70){$p_{01}$}  \put(308,80){.} 
\put(210,80){\vector(3,-1){90}} \put(268,47){$p_{02}$}  \put(308,50){.}
\put(210,50){\vector(3,-1){90}} \put(250,25){$p_{12}$}  \put(308,20){.}

\end{picture}
The lines represent transitions between the 3 different ``error states'' in
which a codeword can be. I have marked only the transitions which will play a
role in the following calculation. In the ``$n$ operations'' step of course
most (100\% to order $\e^0$) codewords will remain error free. The same is
true for the error correction step. I also assume that we have a ``full''
error correction step, which means that to order $\e^0$ all correctable errors
get corrected. In my ``optimized'' implementation of FTEC where most of the
time I measure only one syndrom bit in the $s$-basis and one in the $c$-basis,
we would also have to take into account that $p_{11}>0$, actually $p_{11}=3/7$
or so. It is also a simplification that I put the correctable errors into just
one category. Actually one should make a difference between single bit-flip or
phase-flip errors and a bit-phase-flip error, as they behave differently. I
hope the simplified picture is enough to give an impression of what is going
on. Another approximation is that I assume $p_{01}$ to be small compared to
the amount of correctable errors introduced by the $n$ operations, thus I set
$p_{01}=0$.

The quantity we are looking for, is the probability that a codeword gets an
uncorrectable error. As the diagram shows, such errors happen during the
operations stage as well as during the error correction step. As the $n$
operations are applied bitwise, the error probabilities for different qubits
are independent. This gives $\left({7\atop 2}\right) (n\e)^2$ for the
probability that two errors occurred in different qubits.

With all this we get for the probability of uncorrectable errors per
bitwise operation:

\begin{displaymath} 
\e_1 = \frac{1}{n} \left( \left({7\atop 2}\right) (n\e)^2+7 n \e \cdot
p_{12}+p_{02} \right) 
\end{displaymath}
For fault tolerant error correction we can write in leading order:

\begin{displaymath}
p_{12}=K_{12}~ \e \qquad p_{02}=K_{02}~ \e^2
\end{displaymath}
where the $K$'s are characteristic numbers for the FTEC scheme. With this we
get: 

\begin{displaymath}
\e_1 = \frac{1}{n} ( 21 n^2+7 n \cdot K_{12} +K_{02} ) \e^2
\end{displaymath}
We now have to find the optimal number $n$ of bitwise operations per error
correction step by minimizing this expression. We get:

\begin{displaymath}
n= \sqrt{K_{02}/21}~, \quad \mbox{so}
\quad \e_1=(2 \sqrt{21 K_{02}} +7 K_{12}) \e^2 =K \e^2
\end{displaymath}
Thus the apropriate numbers for characterizing the performance of a FTEC scheme
are $n$ and $K$.

\subsection{Code combinatorics}

The hard part of the calculation is to determine the combinatoric constants
$K_{02}$ and $K_{12}$, which is why eventually I used a computer
simulation. Nevertheless I try to show how in principle one could proceed by
presenting a very crude calculation.

The quantity $K_{02}\cdot\e^2$ is the probability that an initially
undisturbed codeword gets destroyed in an FTEC step due to the occurence of 2
errors. Thus $K_{02}$ is the number of possibilities how two errors in FTEC
(both of probability $\e$) can destroy a codeword. Let's look at $K_{02}$ and
$K_{12}$ for the case where we only try to correct errors in one basis (thus
e.g. only bit flips). Most of the time we will get 0 for the first syndrom bit
and stop there (using here my ``1/3 FTEC'' scheme). Thus both errors would
have to be introduced into the codeword by the syndrom bit measurement. One can
estimate that this doesn't give a big contribution.

It seems that the main contribution to $K_{02}$ comes from the case where one
error causes the first syndrom bit measurement to yield 1 and at the same time
the codeword to get an error.

To estimate these contributions, we should first know the probabilities for
errors in the cat state. Here I take these probabilities from my computer
simulation, although with some patience it could be done by hand. A variant of
my program gives the exact leading term of the Taylor expansion in $\e$. A cat
state has at most 1 phase-flip (pf). I get the following numbers:

\begin{displaymath}
p_{1pf}= 5/3~\e \qquad  p_{1bf}=2/3~\e \qquad  
p_{1pf+1bf}=2/3~\e \qquad  p_{2bf}=39/9~\e^2
\end{displaymath}
From this we next calculate the probabilities of a wrong syndrom bit
measurement and of introducing an error into the codeword. Note that under the
Hadamard transformation the cat phase-flips become bit-flips which give an
erroneous syndrom bit. So the syndrom bit error is:

\begin{displaymath}
p_{sb}=p_{1pf}+p_{1pf+1bf}+4\cdot 2/3~\e=5~\e
\end{displaymath}
where the 3. term is the XOR error. The probability of introducing an error
into the codeword during a syndrom bit measurement is:

\begin{displaymath}
p_{codeword}=p_{1bf}+p_{1pf+1bf}+4~\e \approx 6.3~\e
\end{displaymath}
The first two terms are phase-flips that are transported from ``$H^4$cat'' to
the codeword by the XOR. The 3. term is again from the XOR error. The
probability that both, the syndrom bit and the codeword get wrong is:

\begin{displaymath}
p_{sb+codeword}=p_{1pf+1bf}+4\cdot 2/3~\e \approx 3.3~\e \quad .
\end{displaymath}
Say this happened in the first syndrom bit measurement. Then in the following
3 to 4 such measurements either the codeword gets another error or another
syndrom bit is wrong which may lead to an erroneous error correction attempt
that makes the codeword uncorrectable. With an average number of
$3\frac{1}{2}$ syndrom bit measurements these two probabilities are
$3\frac{1}{2} \cdot p_{sb}$ and $3\frac{1}{2} \cdot p_{codeword}$. Taking only
these contributions we get very roughly:

\begin{displaymath}
K_{02}\approx 3.3 (3.5 \cdot 5+3.5 \cdot 6.3)\approx 130
\end{displaymath}

For $K_{12}$ we take the probability that a second error gets introduced into
the codeword during the 4 syndrom bit measurements. But only a fraction of
about $\frac{2}{3} \cdot \frac{4}{7}$ of the correctable errors will show up
in the first syndrom bit and will thus lead to more measurements, thus we get:

\begin{displaymath}
K_{12}\approx \frac{2}{3} \cdot \frac{4}{7} \cdot 4\cdot 6.3 \approx 9.6
\end{displaymath}

Now we have estimated these values for 1/6 FTEC and there is really no clean
way to obtain from this the corresponding values for a full FTEC step without
knowing more constants. But as we want to get some answer, we simply multiply
both constants with 6 and calculate $K$ from this. We get:

\begin{displaymath}
K\approx 2\sqrt{21\cdot6\cdot130}+7\cdot6\cdot9.6\approx 255+400\approx 650
\end{displaymath}

which confirms the order of magnitude of the threshold given by the computer
simulation.  

\section{Summary}

I claim that the threshold is determined only by the operations needed in a
fault tolerant error correction step. For the 7 bit code these are: XOR,
Hadamard transformation, qubit measurement and preparation of $\ket{0}$. As
long as the Toffoli gate can be applied fault tolerantly to encoded qubits,
its error can eventually be brought down by sufficient concatenation, though
it may always remain much larger than e.g. the error of an XOR.

An improved implementation of FTEC with the 7 bit code allows me to achieve a
threshold of $\e \approx 10^{-3}$ or better for gate errors. 

\section{Outlook}

The present scheme is not thought to be applied directly to a physical quantum
computer, as much more efficient methods may take care of the special error
patterns usually occurring and as it is expected that more space and time
efficient methods exist than concatenation. The result $\e \approx 10^{-3}$
may rather be seen as a guideline about what kind of gate accuracy is
required.

My scheme can certainly be improved further. The following observation won't
help to improve the threshold, but may allow to reduce the number of encoding
levels: If the error $\e$ is sufficiently below the threshold it becomes
worthwile to use longer than 7-qubit codes which can correct more than 1
error. This is particularly true for the higher concatenation levels. Also we
may want to use a code which allows a simpler implementation of the Toffoli
gate (or an equivalent one, see e.g \cite{knill}) and thus a faster reduction
of the Toffoli error.

Eventually a compact unified understanding of fault tolerant quantum computing
may allow more progress as the concatenation scheme is thought to be far from
optimal, especially for space (number of qubits) and time requirements. I
could also imagine that eventually for every quantum algorithm we will try to
find a taylor made fault tolerant version.

\section{Acknowledgements}

I did this work at the Quantum Computation and Quantum Coherence program at
the Institut of Theoretical Physics, University of California, Santa Barbara
and at LANL in November 96. Revisions were made in the following months at
LANL. I would like to thank Manny Knill and Raymond Laflamme for discussions
and especially Adrian Gentle for help with programming in C. I'm supported by
the swiss national science foundation (Schweizerischer Nationalfonds).

\end{document}